\documentclass
[11pt,tightenlines,eqsecnum,floats,aps,amsmath,amssymb,nofootinbib,prd,shownopacs,notitlepage]{revtex4}

\def\be{\begin{equation}}
\def\ee{\end{equation}}
\def\ba{\begin{eqnarray}}
\def\ea{\end{eqnarray}}
\def\bi{\begin{itemize}}
\def\ei{\end{itemize}}
\def\lp{\ell_{\rm Pl}}

\begin{document}

\title{The Overview Chapter in\\
Loop Quantum Gravity: The First 30 Years}
\author{Abhay Ashtekar${}^{1,2}$  and Jorge Pullin${}^{2}$}
\affiliation{
${}^1$ 
Institute for Gravitation and the Cosmos \& Physics Department, The Pennsylvania State University, University Park, PA 16802 U.S.A.
}
\affiliation{
${}^2$
Department of Physics and Astronomy, Louisiana State University, Baton Rouge, LA 70803, U.S.A.
}
\pacs{}
\begin{abstract}
This is the introductory Chapter in the monograph {\it Loop Quantum Gravity: The First 30 Years}, edited by the authors, that was just published in the series ``100 Years of General Relativity''. The 8 invited Chapters that follow provide fresh perspectives on the current status of the field from some of the younger and most active leaders who are currently shaping its development. The purpose of this Chapter is to provide a global overview by bridging the material covered in subsequent Chapters. The goal and scope of the monograph is described in the Preface which can be read by following the Front Matter link at http://www.worldscientific.com/worldscibooks/10.1142/10445.

\end{abstract}
\maketitle

\section{Preamble}
\label{s1}
\begin{quote}

{\it ... a really new field of experience will always lead to
crystallization of a new system of scientific concepts and laws
... when faced with essentially new intellectual challenges, we
continually follow the example of Columbus who possessed the courage
to leave the known world in almost insane hope of finding land beyond
the sea.}

\hspace{8cm}  --Werner Heisenberg 
\end{quote}
\bigskip

This quote from Heisenberg's essay \emph{Changes in the Foundation of Exact Science} succinctly captures the spirit that drives Loop Quantum Gravity (LQG). One leaves behind the \emph{terra firma} of a rigid spacetime continuum in the hope of finding a more supple and richer habitat for physics beyond Einstein. In Columbus' case, while the vision and hope that led to the expedition were indeed almost `insane', he was well aware of the risks. Therefore he embarked on the voyage well prepared, equipped with the most reliable navigational charts and tools then available. Similarly, in LQG one starts with principles of general relativity and quantum mechanics that
are firmly rooted in observations, knowing fully well that one will encounter surprises along the way and habitats at the destination will not look anything like those on these charts. Yet the known charts are essential at the point of departure to ensure that the sails are properly aligned and one does not drift into a fantasy landscape with little relation to the physical world we inhabit.

The five chapters in Part \textbf{2} of this volume describe current status of this voyage and the new habitats it has already led to. One starts with well established general relativity coupled to matter and uses proven tools from quantum mechanics insisting, however, that there be no fields in the background, not even a spacetime metric. This insistence leads one to a rigorous mathematical framework whose conceptual implications are deep \cite{alrev,crbook,ttbook}. Quantum spacetime does not look like a 4-dimensional continuum at all; fundamental excitations of geometry --and hence of gravity-- are polymer like; geometric observables have purely discrete eigenvalues; local curvature in the classical theory is replaced by non-local holonomies of a spin connection; and, quantum dynamics inherits a natural, built-in ultraviolet cutoff. There is no `objective time' to describe quantum dynamics; there are no rigid light cones to formulate causality. Nonetheless, time evolution can be described in detail through \emph{relational} dynamics in the cosmological setting, where familiar causality emerges through \emph{qualitatively new} effective descriptions that are valid all the way to the full Planck regime. Thus, the final landscape is very different from that of general relativity and quantum mechanics, although both provided guiding principles at the point of departure. 

The three Chapters of Part \textbf{3} illustrate the Planck scale `flora and fauna' 
that inhabits this new landscape. The ultraviolet properties of geometry naturally tames the most important singularities of general relativity ---in particular, the big bang is replaced by a quantum bounce. In the very early universe, cosmological perturbations propagate on these regular, bouncing \emph{quantum} geometries, giving rise to effects that are within observational reach. Quantum geometry has also opened a new window on the microscopic degrees of freedom of horizons. Singularity resolution gives rise to a quantum extension of classical spacetimes, creating new paradigms for black hole evaporation in which the evolution is unitary. Finally, several possibilities have been proposed to test LQG ideas in astrophysics and cosmology. They all involve additional assumptions/hypotheses beyond mainstream LQG. Nonetheless, the very fact that relation to observations can be contemplated, sometimes through detailed calculations, provides a measure of the maturity of the subject.

The purpose of this Introduction is twofold: (i) To provide a global overview to aid the beginning researcher navigate through Parts II and III, especially by comparing and contrasting ideas in individual Chapters; and, (ii) Supplement the detailed discussions in these Parts   with a brief discussion of a few general, conceptually important points.  
To keep the bibliography to a manageable size, we will refer only to reviews and monographs (rather than research articles) where further details can be found. \emph{We urge the beginning readers to read this Introduction first, as it spells out the overall viewpoint and motivation that is often taken for granted in individual Chapters.}

\section{The Setting}
\label{s2}

In LQG one adopts the viewpoint that among fundamental forces of Nature,
gravity is special: it is encoded in the very geometry of spacetime.
This is a central feature of GR, a crystallization of the
equivalence principle that lies at the heart of the theory.
Therefore, one argues, it should be incorporated at a fundamental
level in a viable quantum theory. 

The perturbative treatments which dominated the field since the 1960s ignored this aspect of gravity. They assumed that the underlying spacetime can be taken to be a continuum, endowed with a smooth background geometry, and the quantum gravitational field can be treated as any other quantum field on this background. But the resulting perturbation theory around Minkowski spacetime turned out to be non-renormalizable; the strategy failed to achieve the initial goals. The new strategy is to free oneself of the background spacetime continuum that seemed indispensable for formulating and addressing physical questions. In particular, in contrast to approaches developed by particle physicists, one does not begin with quantum matter on a background geometry and then use perturbation theory to incorporate quantum effects of gravity. Matter \emph{and geometry} are both quantum-mechanical at birth. There is often an underlying manifold but no metric, or indeed any other physical fields, in the background.%
\footnote{In 2+1 dimensions, although one begins in a completely analogous fashion, in the final picture one can get rid of the background manifold as well. Thus, the fundamental theory can be formulated combinatorially \cite{aabook}. In 3+1 dimensions, combinatorial descriptions emerge in several approaches to dynamics but one does not yet have a complete theory.}

In classical gravity, Riemannian geometry provides the appropriate mathematical language to formulate the physical, kinematical notions as well as the final dynamical equations. This role is now taken by \textit{quantum} Riemannian geometry. In the classical domain, general relativity stands out as the best available theory of gravity, some of whose predictions have been tested to an amazing degree of accuracy, surpassing even the legendary tests of quantum electrodynamics. Therefore, it is natural to ask: \textit{Does quantum general relativity, coupled to suitable matter exist as a consistent theory non-perturbatively?} There is no implication that such a theory would be the final, complete description of Nature. Nonetheless, this is a fascinating and important open question in its own right.

In particle physics circles the answer to this question is often assumed to be in the negative, not because there is concrete evidence against non-perturbative quantum gravity, but because of the analogy to the theory of weak interactions. There, one first had a 4-point interaction model due to Fermi which works quite well at low energies but which fails to be renormalizable. Progress occurred not by looking for non-perturbative formulations of the Fermi model but by replacing the model by the Glashow-Salam-Weinberg renormalizable theory of electro-weak interactions, in which the 4-point interaction is replaced by $W^\pm$ and $Z$ propagators. Therefore, it is often assumed that perturbative non-renormalizability of quantum general relativity points in a similar direction. However this argument overlooks the crucial fact that, in the case of general relativity, there is a qualitatively new element. Perturbative treatments pre-suppose that the spacetime can be assumed to be a continuum \textit{at all scales of interest to physics under consideration.} This assumption is safe for weak interactions. In the gravitational case, on the other hand, the scale of interest is \emph{the Planck length} $\lp$ and there is no physical basis to pre-suppose that the continuum picture should be valid down to that scale. The failure of the standard perturbative treatments may largely be due to this grossly incorrect assumption and a non-perturbative treatment which correctly incorporates the physical micro-structure of geometry may well be free
of these inconsistencies.

Are there any situations, outside loop quantum gravity, where such physical expectations are borne out in detail mathematically? The answer is in the affirmative. There exist quantum field theories (such as the Gross-Neveu model in three dimensions) \index{Gross-Neveau model} in which the standard perturbation expansion is not renormalizable although the theory is \emph{exactly soluble} \cite{jf}! Failure of the standard perturbation expansion can occur because one insists on perturbing around the trivial, Gaussian point rather than the more physical, non-trivial fixed point of the renormalization group (RG) flow. Interestingly, thanks to developments in the Asymptotic Safety program \index{asymptotic safety}  there is now growing evidence that situation may be similar in quantum general relativity \cite{reuterrev}. Although there are some basic differences \cite{arr} between LQG and the Asymptotic Safety program, these results provide concrete support to the idea that non-perturbative treatments of quantum general relativity can lead to an ultraviolet regular theory.%
\footnote{In the Asymptotic Safety program, spacetime geometry in the Planck regime is effectively 2-dimensional as in LQG and the 4-dimensional continuum arises only in the low energy limit \cite{reuterrev}.} 

However, even if the LQG program could be carried out to completion, there is no a priori reason to assume that it would be the `final' theory of all known physics. In particular, as is the case with classical general relativity, while requirements of background independence and general covariance do restrict the form of interactions between gravity and matter fields and among matter fields themselves, the theory would not have a built-in principle which \textit{determines} these interactions. Put differently, such a theory may not be a satisfactory candidate for unification of all known forces. However, just as general relativity has had powerful implications in spite of this limitation in the classical domain, LQG should have qualitatively new predictions, pushing further the existing frontiers of physics. Indeed, unification does not appear to be an essential criterion for usefulness of a theory even in other interactions. QCD, for example, is a powerful theory even though it does not unify strong interactions with electro-weak ones. Furthermore, the fact that we do not yet have a viable candidate for  grand unified theory does not make QCD any less useful. Finally, as the three Chapters in Part \textbf{3} illustrate, LQG has already made interesting predictions for quantum physics of black holes and the very early universe, some of which are detailed and make direct contact with observations.

\section{Quantum Riemannian Geometry}
\label{s3}

Since the basic dynamical variable in general relativity is the spacetime metric, Wheeler advocated the view that we should regard it as \emph{geometrodynamics},\, a dynamical theory of 3-metrics $q_{ab}$ that constitute the configuration variable. For the three other basic forces of Nature, on the other hand, the dynamical variable is a connection 1-form that takes values in the Lie algebra of the appropriate internal group. In QED the connection enables one to parallel transport electrons and positrons while in QCD it serves as the vehicle to parallel transport quarks. Now the configuration variable is a spatial connections $A_{a}^{i}$; we have theories of \emph{connection-dynamics.} Weinberg, in particular, has emphasized that this difference has driven a `wedge between general relativity and the theory of elementary particles' \cite{weinberg}. 

As described in {\it Chapter 1}, the starting point in LQG is a reformulation of general relativity as a dynamical theory of spin connections \cite{aabook}. We now know that the idea can be traced back to Einstein and Schr\"odinger who, among others, had recast general relativity as a theory of connections already in the fifties. (For a brief account of this fascinating history, see \cite{aa1}.) However, they used the `Levi-Civita connection' that features in the parallel transport of vectors and found that the theory becomes rather complicated. The situation is very different with self-dual (or anti-self-dual) spin connections. For example, the dynamical evolution dictated by Einstein's equations can now be visualized simply as a \emph{geodesic motion} on the `superspace' of spin-connections (with respect to a natural metric extracted from the constraint equations). Furthermore, the (anti-)self-dual connections have a direct physical interpretation: they are the vehicles used to parallel transport spinors with definite helicities of the standard model. With this formulation of general relativity, Weinberg's `wedge' disappears. In particular, phase-space of general relativity is now the same as that of gauge theories of the other three forces of Nature \cite{aabook,jbbook}. However, in the Lorentzian signature, the (anti-)self-dual connections are complex-valued and, so far, this fact has been a road-block in the construction of a rigorous  mathematical framework in the passage to the quantum theory.%
\footnote{At a formal level in which one often works in quantum field theory (QFT), one can carry out calculations with complex connections. 
The road block refers to rigorous mathematical constructions. Specifically, it has not been possible to introduce measures and develop differential geometry on the infinite dimensional space of generalized connections that serve as the quantum configuration space. However, as is the standard practice in constructive QFT, it may be possible to work in the Riemannian signature where (anti-)self-dual connections are real and then pass to the Lorentzian theory through a \emph{generalized} Wick transform that does exist \cite{ttbook}. This is an opportunity for future research.}
Therefore, the strategy is to pass to real connection variables by performing a canonical transformation \cite{alrev,crbook,ttbook}. The canonical transformation introduces a real, dimensionless constant $\gamma$, referred to as the Barbero-Immirzi parameter; \index{Barbero-Immirzi parameter} the (anti-) self-dual Hamiltonian framework is recovered by formally setting $\gamma = \pm i$. 

For real connection variables  the `internal gauge group' reduces to SU(2), which is compact. Therefore, as explained in {\it Chapter 1,} it is possible to introduce integral and differential calculus on the infinite-dimensional space of (generalized) connections rigorously without having to introduce background geometrical fields, such as a metric. Since this space serves as the quantum configuration space, one can introduce a Hilbert space of square integrable functions and physically interesting self-adjoint operators thereon. This setup then serves as the kinematical framework in LQG. %
\footnote{In rigorous QFT in Minkowski space, while the set of smooth fields that constitute the classical configuration space is topologically dense in the quantum configuration space on which wave functions live, the quantum measure is concentrated on genuinely distributional fields. In LQG the situation is similar. Smooth connections are topologically dense in the space of generalized ones but contained in a set of measure zero. This level of rigor is essential if the kinematical framework is to serve as the point of departure for quantum dynamics. It was never attained in geometrodynamics ---the Wheeler-DeWitt framework has remained entirely formal.} 

The most important features of this background independent framework are the following. {\it First,} just as one has the von-Neumann uniqueness theorem in quantum mechanics of a finite number of degrees of freedom \index{Stone-von Neumann theorem} that singles out the standard Schr\"{o}dinger representation, there are uniqueness theorems that single out this kinematical framework. This is a highly non-trivial result because the system has an infinite number of degrees of freedom, made possible only because the requirement of background independence is very stringent. {\it Second,} in this framework, geometric operators describing the quantum Riemannian geometry have purely discrete eigenvalues. This is a striking and unforeseen outcome, given that the point of departure is standard general relativity. For example, this discreteness is \emph{not} shared by quantum geometrodynamics (i.e., by the Wheeler-DeWitt theory). 

There is a basis in the LQG kinematical Hilbert space which is well-suited to analyze properties of these geometric operators: \emph{spin network basis} \cite{alrev,crbook,ttbook}. The notion of spin networks was introduced by Penrose already in 1971 in a combinatorial approach to the Euclidean 3-geometry. This notion was generalized in LQG.  Now spin networks are labelled by graphs in which any number of links can meet at nodes and both inks and nodes carry certain information ---called `decorations' or `colors'  \index{spin networks/spin-nets} This kinematics brings out the precise sense in which the fundamental excitations of (spatial) geometry in LQG are 1-dimensional, polymer like. Consider for simplicity  graphs which have only 4-valent nodes ---i.e. in which precisely 4 links meet at each node. Then, one can introduce a simplicial decomposition of the 3-manifold which is dual to the graph: Each node of the graph is contained in a single tetrahedron and each link of the graph associated with that node intersects precisely one face of that tetrahedron. The `decoration' on the links assigns specific quantized areas to faces of tetrahedra and the `decoration' at the node determines the volume of that tetrahedron. Thus, each topological tetrahedron can be regarded as an `atom of space', characterized by the decorations. \index{atom of space} The volumes of tetrahedra and areas of its faces endow them with geometrical properties. But these geometries are not induced by smooth metrics on the 3-manifold.  One thinks of the familiar continuum Riemannian geometries as `emergent', arising from a coarse graining of the fundamental quantum geometry.

Of particular interest is the smallest non-zero eigenvalue of the area operator, called the \emph{area gap} whose value in Planck units is denoted by $\Delta$. \index{area gap} It turns out that $\Delta$ depends linearly on the Barbero-Immirzi parameter: $\Delta = 4 \sqrt{3}\pi \gamma$. Therefore one can trade one for the other. Conceptually, it is appropriate to regard $\gamma$ as a `mathematical parameter' that features in the transition from classical to quantum theory, and $\Delta$ as the `physical parameter' that sets the scale at which quantum geometry effects become important. Thus, from the perspective of the final quantum theory $\Delta$ is the fundamental physical parameter. For example, in Loop Quantum Cosmology (LQC) of homogeneous isotropic models, energy density has a maximum value given by $\rho_{\rm max} = (18\pi/\Delta^{3})\, \rho_{\rm Pl}$ and as $\Delta \to 0$, i.e., as we ignore quantum geometry effects, we recover the classical result $\rho_{\rm max} \to \infty$. This is completely analogous to the situation for the energy spectrum of the Hydrogen atom in quantum mechanics: the ground state energy is given by $E_{0} = - (me^{4}/2 \hbar^{2})$ and $E_{0} \to -\infty$, the classical value, as $\hbar \to 0$. More generally, one should formulate physical questions in terms of $\Delta$, replacing $\gamma$ by $ \Delta/(4\sqrt{3}\pi)$ in expressions of interest. In the older LQC literature, some confusion arose because one took the limit $\Delta \to 0$, keeping $\gamma$ fixed in expressions that also involved $\gamma$. Similarly, the discussion of black hole entropy becomes significantly clearer if everything is formulated in terms of $\Delta$. In particular, since it is physically clear that the microcanonical calculations of entropy using Planck scale quantum geometry should have a dependence on the value of the area gap, one is not tempted to find arguments to make it independent of $\gamma$.%
\footnote{In semi-classical considerations, arguments are restricted to states representing `near horizon, classical geometries' and one calculates entropy using canonical and micro-canonical ensembles. Then the final result can be insensitive to the area gap $\Delta$.} 

The quantum geometry framework is described in detail in {\it Chapter 1} and lies at the foundation of issues discussed in later Chapters.

\section{Non-perturbative, Background-independent Dynamics}
\label{s4}

{\it Chapters 2-5} summarize the current status of dynamics in full LQG from both the Hamiltonian and path integral perspectives. 

In the early years of LQG the primary focus was on considering general relativity as a constrained Hamiltonian theory --but not in terms of 3-metrics and extrinsic curvature, but rather in terms of spin connections and their canonically conjugate momenta, the spatial triads (with density weight 1). Because the phase space is the same as in an SU(2) Yang-Mills theory, one could import into gravity the well-developed techniques from gauge theories. However, unlike the familiar Yang-Mills theory in Minkowski spacetime, now there is no metric or any other field in the background. In particular expressions of general relativity constraints involve only the \emph{dynamical} phase space variables. Since there are no fiducial geometric structures, and coordinates themselves can be rescaled arbitrarily without affecting underlying physics, techniques used in Minkowskian quantum field theories to regulate products of field operators are no longer useful. Therefore a number of novel and astute techniques had to be developed to construct the physical sector of the theory by imposing quantum constraints a la Dirac in the background independent kinematical framework of {\it Chapter 1}. (In the Hamiltonian  framework this is equivalent to solving quantum Einstein's equations.) These developments are discussed in {\it Chapter 2}. However, as discussed there, to make the constraint operators well-defined in the rigorous setting provided by the kinematical framework, one had to introduce a number of auxiliary structures. This is not surprising by itself; such constructions are also needed to regulate products of operators in Minkowskian quantum field theories. However, in LQG the \emph{final} physical sector of the theory depended on the scheme chosen and the physical meaning of these differences remained opaque. 

What was needed was a principle to streamline the calculations and reduce the available freedom. Perhaps the most natural and most attractive of these principles is to demand that the quantum constraint algebra should be closed not only on quantum states that satisfy all or some of the constraints, but on the full kinematical state space. The regularization strategies used in the 1990s could not shed light light on this issue of `off-shell closure' of quantum constraints. Over the past five years or so, this idea is being systematically implemented in models with increasing complexity. The task is technically difficult. A key idea is to use the fact that, in connection dynamics, the difficult Hamiltonian constraint can be regarded as a diffeomorphism constraint in which the shift fields themselves have certain specific dependence on the phase space variables i.e., are so called `q-number' quantities. (This simplification does not occur in geometrodynamics, where the two constraints have entirely different forms.)

Therefore effort is directed to constructing quantum operators that generate infinitesimal (generalized) diffeomorphisms. (In the early LQG  works one only had operators implementing \emph{finite} and `c-number' diffeomorphisms.) The strategy works in simpler models such as parametrized field theories, \index{parametrized field theory} which were invented as simpler models that mimic features associated with background independence of general relativity. In these theories, the new LQG techniques have enabled one to overcome some long standing obstacles and construct a satisfactory quantum theory with fundamental discreteness as well as covariance. These ideas have also been successfully applied to certain models that arise from simplifications of general relativity. In these systems, the freedom in quantization is neatly streamlined by the requirement of `off-shell' closure of quantum constraints. These systematic advances, summarized in {\it Chapter 2}, have opened concrete directions to complete the Dirac program in the framework of connection dynamics. \index{Dirac quantization} \\

{\it Chapter 3} introduces the basics of spin foams,\index{spin foams} the sum over histories approach which has been a primary focus of recent work on quantum dynamics \cite{perez,crfv}. Recall that in his original derivation of path integrals, Feynman began with the expressions of transition amplitudes in Schr\"odinger quantum mechanics and \emph{reformulated} them as an integral over all kinematically allowed paths \cite{fh}. In background independent theories, on the other hand, we have a constrained Hamiltonian framework. As we saw above, in the Dirac program, physical states simply solve the quantum constraints and one has to tease out dynamics, e.g., by introducing a relational time variable. Therefore,  on formally mimicking the Feynman procedure starting from the Hamiltonian framework, one finds that the analog of the transition amplitude is an \emph{extraction amplitude}. This is a Green's function that extracts from `incoming' (or `outgoing') kinematical states, solutions to quantum constraints and also provides the physical inner product between them. Thus, path integrals provide an alternate, covariant avenue to construct the physical Hilbert space of the theory. If the theory can be deparametrized, it inherits a relational time variable and then the extraction amplitude can be re-interpreted as a transition amplitude with respect to that time.%
\footnote{In LQC these steps have been carried out rigorously in the  Friedmann-Lema\^{i}tre-Robertson-Walker (FLRW) as well as Bianchi I models. That is, one can begin with the well-defined Hamiltonian quantum theory and arrive at the covariant cosmological spin foam, obtain the exact Green's function for the extraction amplitude, show that it admits a `vertex expansion' that is convergent, and interpret the extraction amplitude as a transition amplitude by an appropriate deparameterization   \cite{asrev}.}  
Irrespective of whether this is possible, the basic object that encodes quantum dynamics is the extraction amplitude.

In heuristic treatments of sum over histories, kinematical paths are generally represented by smooth classical fields. However, in rigorous QFT it is well known that these paths are contained in a set of measure zero; the measure is concentrated on genuinely distributional fields. Situation is very similar in LQG: the measure is concentrated on generalized connections rather than smooth ones (see footnote (d)). Consequently in spin foams the sum is over \emph{quantum} spacetime geometries rather than smooth metrics. These are represented by `decorated' 2-complexes that can be heuristically thought of as `time evolution' of spin networks. Quantum geometries associated with a given 2-complex can be regarded as paths that interpolate between given `incoming' and `outgoing' spin networks, the `decorations' providing specific spacetime quantum geometries which are described in some detail in {\it Chapter 3}.

In any given 2-complex, (zero-dimensional) nodes of the `incoming' spin network `evolve' into  (one-dimensional) edges, but every now and then a vertex is created characterizing a `non-trivial happening'. There is no time, yet `happenings' are objectively coded in each quantum history. Each `decorated' 2-complex carries a fixed number of such happenings but can represent many different quantum geometries depending on the choice of decorations. In \emph{any} choice of decorations, areas of 2-dimensional faces are quantized and there is a minimum non-zero area ---the area gap. Thus, properties of the underlying quantum geometry provide a natural ultraviolet cut off. To carry out the path integral, one has to assign amplitudes to the faces and vertices of the 2-complex. The non-trivial part turns out to be the specification of the vertex amplitude. The first concrete prescription was given in the Barrett-Crane model which opened up the field of spin foams. However, later attempts to calculate the graviton propagator in Minkowski spacetime starting from non-perturbative spin foams revealed some important limitations of this model. It was replaced by the Engle--Pereira-Rovelli-Livine (EPRL) and (the closely related) Freidel-Krasnov (FK) models.\index{Engle-Pereira-Rovelli-Livine (EPRL) model}\index{Freidel-Krasnov (FK) model} In these models, the built-in, natural cut off makes the integral over `decorations' ultraviolet finite in any given 2-complex. It is unlikely that there is an analogous infrared finiteness in general relativity with zero cosmological constant. However, in presence of a positive cosmological constant, one is naturally led to replace the SU(2) group of internal rotations (of spinors) with its its quantum analog ${\rm SU(2)_{q}}$ and the amplitudes have been shown to be infrared finite as well.

In full LQG, of course, one must go beyond a single 2-complex and consider `all possible' 2-complexes interpolating between the two spin networks, allowing for an arbitrary number of vertices. {\it Chapters 4 and 5} summarize two complementary but different approaches to fulfill this task. \\

{\it Chapter 4} presents a more general view of sum over histories which, 
at first, seems very different from spin foams but in fact first arose as a `generalized Fourier transform' from spin foam models. The point of departure is neither a Hamiltonian formulation of general relativity as in the canonical approach, nor a `constrained topological theory' with which one starts in spin foams. Not only are there no background fields such as a metric but there is no spacetime manifold at a fundamental level. The underlying idea is that gravity is \emph{to emerge} from a more fundamental theory based on abstract structures that, to begin with, have nothing to do with spacetime geometry. Drawing inspiration from the matrix models (in 2 spacetime dimensions) and especially the `Boulatov-model' (in 3 spacetime dimensions) the fundamental object is a QFT but  formulated on a \emph{group manifold, rather than spacetime}. Therefore the framework is aptly called  `Group Field Theory' (GFT).\index{group field theory} As in familiar field theories, the Lagrangian has a free and an interaction term, with a coupling constant ${\lambda}$. Even though the point of departure appears to be so different, one can again use the LQG kinematics and represent the `in' and the `out' states by spin networks as in spin foams.%
\footnote{Each spin network captures only a finite number of degrees of freedom of the quantum gravitational field which can be interpreted as a `twisted geometry'. In GFT, one often says that they represent a `first quantized' theory. The full GFT has operators that create spin network states and is therefore regarded as a `second quantized' theory. This terminology is \emph{not} used outside GFT and `second quantization' should not be interpreted as going beyond LQG:\, LQG has the same underlying mathematical structures.} 
Remarkably, for a certain choice of the Lagrangian, the $n$-th term in the perturbation expansion --i.e., the coefficient of $\lambda^{n}$ -- is the same as the contribution to the extraction amplitude obtained by fixing a 2-complex with precisely $n$ vertices, and summing over the decorations in the EPRL model. But because it arises in a standard perturbation expansion --albeit on a group manifold, not spacetime-- one can now borrow techniques from standard QFT. This is especially important in order to (i) go beyond a single 2-complex (i.e., a fixed number of vertices); and (ii) analyze potentially distinct phases, as in standard QFT. 

In this summary we have presented GFT from the perspective of Hamiltonian and spin foam LQG (although, as explained in {\it Chapter 4}, there are also some technical differences). However, GFT offers greater generality. For example, GFT naturally suggests quantum LQG dynamics of a more general, `grandcanonical' type in which the number of vertices is allowed to vary. It also opens avenues to discuss the `continuum limit' using ideas in the spirit of the thermodynamic limit in quantum field theories. In addition, it suggests that even if one begins with GFT actions corresponding to, say, the EPRL model, under the RG flow quantum dynamics will generate many more terms at different scales, providing new scale dependent physics. Finally, the framework is so general that it may even allow a departure from a fundamental tenet of the rest of LQG that there are no degrees of freedom beyond the Planck scale. Thus, like QFT in Minkowski spacetime, GFT offers a general paradigm, rather than a physical theory. To specify a physical theory, one has to choose a set of fields --now on a group manifold-- and fix interactions between them. Hence its scope differs from the more focused approaches to quantum dynamics discussed in {\it Chapters 2 and 3}. In recent years the emphasis has been on exploiting the generality it offers by applying the ideas to specific models. As in {\it Chapter 2}, it has been successfully applied to simpler systems. 
The state of the art can be summarized by saying that GFT has opened a number of new avenues that have the potential to resolve the key open issues in 4-dimensional spin foams.\\

Finally, in {\it Chapter 5} we return to the important open issue of the continuum limit in spin foams, which is now taken using a generalization of the standard RG flow induced by refinements in the spacetime manifold in the spirit of lattice QCD, rather than through a field theory on a group manifold. Here, one uses methods from lattice QFT, functional analysis and tensor networks, rather than perturbative expansions in coupling constants together with techniques from non-perturbative QFTs, used in GFT. More precisely, the physical states --i.e., solutions of the quantum constraint-- are to be  constructed by taking the refinement limit in spacetime.

Let us begin with the kinematical setup of {\it Chapter 1}. The strategy employed there is to first define structures, such as the scalar product between states and action of geometric and holonomy operators on them, using the finite degrees of freedom that are captured in a single graph. These structures then naturally extend to the full state space in the continuum that captures all the infinitely many degrees of freedom \emph{provided} they satisfy stringent consistency conditions as one coarse grains or refines graphs by adding new nodes and links. These are the so-called \emph{cylindrical consistency conditions} \cite{alrev}. Quantum geometry has been successfully constructed in the kinematical setting precisely because these cylindrical consistency requirements were met. The idea now is to promote these consistency requirements \emph{to dynamics}, using spin foams of {\it Chapter 3}.

One can start with a fixed simplicial decomposition of the spacetime manifold together with its dual 2-complex. The 2-complex induces spin network states on the initial and final 3-manifolds which can be regarded as the `in' and `out' quantum states of geometry (or, more precisely, simply kinematic states out of which one wishes to extract physical ones). As noted above, a spin foam model provides a `transition amplitude' between them (or, more precisely, a Green's function that extracts a physical states from the two given kinematical ones). The idea is that one should consider only those refinements (and coarse grainings) of the simplicial decomposition --and hence of the dual 2-complex-- that make the procedure cylindrically consistent, so that physics at different `scales' is appropriately related to constitute a coherent scheme. This is the sense in which the notion of cylindrical consistency is to be  promoted from the kinematical to the dynamical setting. A refinement procedure that meets the consistency conditions would then provide a generalization of the RG flow ideas to a setting in which there is no background metric to define the scale. The question of whether such a refinement limit exists is similar to the asymptotic safety conjecture that an ultraviolet fixed point exists. In the same spirit as asymptotic safety, once the program is developed beyond pure gravity, the hope is that cylindrical consistency will severely constrain matter couplings since it is a stringent requirement.

{\it Chapter 5} explains these consistency conditions and also the inductive limit  that is to provide the final quantum dynamics in the continuum limit through an admissible refinement. Again, as in {\it Chapters 2 and 4}, the procedure has been successfully applied to simpler models, now involving decorated tensor networks and  an iterative procedure. The general viewpoint in LQG is that the so called Ashtekar-Lewandowski (AL) representation (of the fundamental LQG quantum algebra) underlying the kinematic setup correctly captures the essential features of quantum geometry at Planck scale. But by loosening the requirements of that led to the uniqueness of this representation, one can construct a `dual' description, called the BF representation. General arguments have been put forward to suggest that it may be more directly useful for describing the phase of the theory  containing macroscopic, continuum geometries.\index{Ashtekar-Lewandowski representation}\index{BF representation} Each of these representations has a vacuum state and the two vacua are very different from one another. An important feature of the overall strategy is that the truncation scheme is to be determined by dynamics. `Coarse states' will have few excitations while `fine states' will have many excitations with respect to a vacuum that is also determined by dynamics.

This approach to continuum limit has several attractive features. First, the procedure brings out a close relation between the the continuum quantum dynamics and spacetime diffeomorphism symmetry in systems with (auxiliary) discrete structures, reflecting the intuitive idea that the diffeomorphism symmetry allows one to refine or coarse grain any region. Second, it provides a background independent analog of a `complete renormalization trajectory' through the notion of cylindrically consistent amplitudes. Finally, the procedure already has the necessary ingredients in place to lead to the running of coupling constants, once the system is extended to allow for matter sources. As in {\it Chapters 2 and 4}, this is an ongoing program; now the open question is whether cylindrically consistent amplitudes exist in full 4-dimensional LQG.

\section{Applications}
\label{s5}

While important issues remain in full LQG, the basic underlying ideas have been successfully applied to two physical sectors of full gravity: black holes and the very early universe. Part {\bf 3} of this volume summarize these advances.

Cosmic microwave background (CMB) observations have established  that the early universe is spatially homogeneous and isotropic to one part in $10^{5}$. Therefore the current paradigms of the early universe assume that spacetime geometry is well described by a Friedmann-Lema\^{i}tre-Robertson-Walker (FLRW) geometry, together with first order perturbations prior to the CMB epoch. Although we do not yet have a definitive paradigm, the inflationary scenario has emerged as the leading candidate. In particular, it has been very successful in accounting for the one part in $10^{5}$ inhomogeneities observed in the CMB. Known physics and astrophysics shows that these inhomogeneities serve as seeds for formation of the large scale structure we see in the universe. Thus, inflation pushes the issue of the origin of the observed large scale structure further back from the CMB epoch ---in fact to \emph{very} early times, when the spacetime curvature was some $10^{62}$ times that at the surface of a solar mass black hole! While this is truly impressive, from the viewpoint of quantum gravity this epoch lies in the classical general relativity regime since the curvature is still some $10^{-11}$ times the Planck curvature. That is why it is consistent --as is done in all current paradigms of the early universe scenario-- to describe the universe using a classical FLRW background and represent the cosmological perturbations by quantum fields propagating on it. However, this strategy is inadequate if one wishes to go to still earlier times and describe what happened when the matter density and curvature were of Planck scale. For this, one needs a quantum theory of gravity. The inadequacy of standard inflation is brought out by two facts: (i) the big-bang singularity persists in this theory, and all physics comes to a halt there; and (ii) quantum field theory on FLRW backgrounds, used to describe the dynamics of cosmological perturbations, becomes inadequate because even modes that can be observed in the CMB acquire trans-Planckian frequencies in the early epoch. Thus, a challenge to any candidate quantum gravity theory is to provide a completion of the inflationary scenario over the $11$ orders of magnitude in matter density and curvature  that separate it from the Planck scale and successfully address these issues. As {\it Chapter 6} describes, in LQG there have been remarkable advances in this direction.\\

Let us begin with the first issue ---that of the resolution of the big-bang singularity in the background spacetime. In LQG cosmological singularities are resolved in all models that have been studied so far. These include the flat and closed FLRW models with and without a cosmological constant (of either sign); the anisotropic Bianchi models that contain non-linear gravitational waves; and the inhomogeneous Gowdy models\index{Bianchi models}\index{FLRW spacetimes}\index{Gowdy models} which also contain  non-linear gravitational waves \cite{asrev}. The resolution does not come about by introducing matter that violates energy conditions or by some fine tuning. The origin of the mechanism can be traced back directly to the underlying quantum geometry --particularly the emergence of the area gap--  described in {\it Chapter 1}. Quantum geometry creates a brand new repulsive force. It is completely negligible already at the onset of inflation and thereafter. However if we evolve back in time, it grows rapidly, overwhelms the classical attraction in the Planck regime, and causes the universe to bounce. Thus the big bang is replaced by a quantum bounce. More generally, the following picture succinctly summarizes the salient features of quantum dynamics in cosmological models: anytime a curvature invariant starts to grow in general relativity signaling approach to a singularity, the repulsive force grows to dilute it, preventing its formation.  

What is behind this singularity resolution? In Loop Quantum Cosmology (LQC) one applies the LQG techniques to the cosmological setting described above. It turns out that even after standard gauge fixing there is still some residual diffeomorphism freedom in cosmological models. The requirement of covariance under this freedom --i.e., background dependence in the cosmological context-- again leads to a unique representation of the fundamental quantum algebra. As in the AL (or BF) representation for full LQG, the connection operator is not well defined in LQC; only its exponential, the holonomy, is a well defined (unitary) operator. In this LQC representation of quantum states, then, (as in full LQG) one has to express the curvature term in the Hamiltonian constraint in terms of holonomies, now around a loop which encloses the minimum possible physical area defined by the quantum state. As a result the quantum Hamiltonian constraint depends on the area gap and the operator reduces to the Wheeler DeWitt operator of geometrodynamics only in the limit in which the area gap goes to zero. At a fundamental level, there is no time. But in presence of suitable matter, one can deparameterize the theory and interpret the Hamiltonian constraint as an evolution equation in a relational time variable provided, e.g., by matter. One can show that physical (Dirac) observables --such as the energy density or curvature at a given instant of relational time-- that diverge at the big bang in general relativity have a finite upper bound on the entire space of physical states. Thus the singularity is resolved in a very direct, physical sense: There are simply \emph{no} states in the physical Hilbert space in which matter density diverges. This resolution has been analyzed using Hamiltonian and path integral methods and has also been discussed in the `decoherent histories' framework. {\it Chapter 6} summarizes most of these results and provides references for topics that could not be covered.

Thus, in LQC, the classical FLRW backgrounds $(a(t), \phi(t))$ used in the inflationary scenario are replaced by a wave function $\Psi(a, \phi)$ in the physical Hilbert space. Quantum fields representing the scalar (curvature) and tensor perturbations now propagate on the quantum FLRW geometry $\Psi(a, \phi)$. Over the last several years, quantum field theory on classical FLRW spacetimes was systematically generalized to quantum field theory on these \emph{quantum} FLRW spacetimes. This generalization enables one to face the `trans-Planckian issues' squarely and has therefore been used to analyze dynamics of the scalar and tensor modes through the Planck regime, all the way to the quantum bounce. This analysis, as well as other methods described in {\it Chapter 8}, have led to interesting phenomenological predictions that can be tested against observations. 

Of particular interest is an unforeseen interplay between the ultraviolet and the infrared. Quantum geometry effects that tame the singularity provide a new ultraviolet LQC length scale $\ell_{\rm LQC}$, the minimum curvature radius ---corresponding to the maximum scalar curvature at the bounce.  Modes of perturbations whose physical wavelength $\lambda_{\rm phys}$ is \emph{larger} than $\ell_{\rm LQC}$ experience curvature during their evolution in the Planck regime near the bounce and are excited. These turn out to have the longest wavelength among the modes seen in the CMB observations. Thus, dynamics in the Planck regime can leave imprints at the largest angular scales in the sky. Detailed calculations have been performed to use this effect to account for the anomalies that the PLANCK and WMAP teams have found at the largest angular scales in CMB. Furthermore, predictions have been made for other correlation functions that should soon be reported by the PLANCK team for the largest angular scales. Thus, LQC has been a fertile ground within LQG. It has led to a concrete and detailed realization of several key underlying ideas  ---use of relational time, construction of a complete set of Dirac observables, relation between the canonical framework and spin foams, and the hope that the vertex expansion in spin foams is convergent. At the same time, it has brought quantum gravity from lofty heights of mathematical physics to concrete, observational issues in phenomenology.\\

{\it Chapter 7} summarizes results on the black hole sector of LQG. Just as the early universe provides us with possibly the best opportunity of directly observing quantum gravity effects, black holes provide us with possibly the best arena to test quantum gravity theories at a conceptual level. Specifically any viable quantum gravity has to address two issues in the black hole sector. The \emph{first} arises from black hole thermodynamics. Einstein equations within classical general relativity inform us that black holes obey certain laws: the zeroth refers to equilibrium configurations, the first to transition from an equilibrium state to a nearby one, and the second to full dynamical situations. Remarkably, these laws become the zeroth, the first and the second laws of thermodynamics if one identifies a multiple of the surface gravity $\kappa$ of the horizon with temperature $T$, and ($1/(8\pi\, G)$ times the) reciprocal multiple of the horizon area $A_{\rm H}$ with (Bekenstein-Hawking) entropy $S_{\rm BH}$. However, purely from dimensional grounds one finds that the multiple must have the same dimensions as $\hbar$, bringing in quantum mechanics in a totally unforeseen fashion.  Subsequent calculation, using quantum field theory in the Schwarzschild black hole spacetime, established the Hawking effect: Black holes evaporate quantum mechanically and at late times the outgoing state is extremely well approximated by the thermal radiation from a black body at the (Hawking) temperature $T_{H} = \kappa\hbar/2\pi$. These discoveries, made over 4 decades ago, made it clear that black holes hold a key to bring together the three pillars of physics ---general relativity, thermodynamics and quantum mechanics. They imply that a solar mass black hole has $\exp 10^{77}$ microstates, an enormous number even on statistical mechanics standards. One is thus immediately led the question: Can we account for the enormous horizon entropy $S_{\rm BH}$ using more fundamental statistical mechanical considerations? The \emph{second} issue concerns the dynamics of the evaporation process. What happens in a self-consistent treatment when one includes back reaction so that, by energy conservation, the black hole mass decreases? Does the horizon evaporate completely? The black hole may have been formed by sending in a variety of, say, pure states. Are we left with particles in a thermal state at the end of evaporation? If so, the evaporation process would not be unitary and information would be lost. Even after four decades, there are no clear answers to this second set of questions in any approach to quantum gravity. Therefore {\it Chapter 7} focuses only to first set of issues in this section. \index{Hawking radiation}

Event horizons that one normally associates with black holes are {\it extremely} global as well as teleological notions. For example, they can form and grow in a flat, Minkowskian region of a spacetime. Similarly, they can be gotten rid of simply by changing spacetime geometry in a tiny neighborhood of the singularity. Furthermore, it is not unlikely that the spacetime describing an evaporating black hole has an event horizon. Calculations with 2-dimensional black holes that include the back reaction and solve for geometry using a mean field approximation strongly indicate that the full spacetime will not have an event horizon. Therefore, as explained in detail in {\it Chapter 7}, in LQG one uses the quasi-local notion of a (weakly) \emph{isolated} horizon for which the zeroth and the first laws of black hole mechanics do hold \cite{akrev}. An additional advantage is that one can incorporate \emph{both} the black hole and the cosmological horizons in one swoop, and one does \emph{not} have to restrict oneself to near extremal black holes.%
\footnote{To obtain the second law, one needs \emph{dynamical} horizons which are also quasi-local notions. While event horizons are null hypersurfaces, dynamical horizons are space-like when the black hole is growing during collapse and time-like when it is evaporating. Unlike event horizons, they are \emph{not} one way membranes. This fact removes considerable confusion in the literature on the evaporation process. There is strong indication that dynamical horizons do exist for evaporating black holes. (When matter is neither falling into the black hole nor leaving it, the dynamical horizon becomes null and an isolated horizon.) \cite{akrev}} 

In LQG, one focuses on the quantum geometry of isolated horizons. The `quanta of geometry' provide the microstates of the quantum horizon that are used in a statistical mechanical entropy calculation. The subject is mathematically rich. It involves quantum groups, Chern-Simons theory on punctured spheres, mapping class groups and sophisticated techniques from number theory. Detailed calculations show that the number of microstates has interesting properties, especially for small black holes. For large black holes, the entropy $S_{\rm BH}$ is indeed proportional to the horizon area $A_{\rm H}$ but this coefficient is inversely proportional the area gap $\Delta$ of LQG. This is just as one would expect on general grounds. So, as the area gap goes to zero, i.e., we ignore quantum geometry effects, the entropy becomes infinite. However, as a result, the Bekenstein-Hawking formula $S_{\rm BH} = A_{\rm H}/4G\hbar$ is recovered only for a certain value of the area gap $\Delta$. Put differently, by demanding that LQG have the correct semi-classical limit to leading order, one can fix the value of the area gap, and hence remove the a priori 1-parameter ambiguity in the kinematical setup of LQG. This general picture is well-established. However, as the discussion in {\it Chapter 7}\, shows, there are ambiguities in the precise definition of what constitutes a quantum state of the horizon geometry, primarily because the surface states of the horizon quantum geometry are entangled with states of the bulk quantum geometry. Therefore, at a conceptual level, the subject is not fully settled. However, the ambiguities are all very small to make a difference in practice: The area gap varies only between $5.16$ and $5.96$ (in Planck units).

These considerations refer to quantum geometry and Planck scale eigenvalues of the area operator. In recent years, there has been a significant advance in relating LQG to the vast literature on black hole thermodynamics based on semi-classical considerations. The main idea is to consider the near horizon geometry corresponding to that of a stationary black hole solution and shift the perspective to that of a suitable family of near horizon stationary observers. In such geometries, one can relate the energy $E$, as seen by stationary observers at a distance $\ell$ from the horizon, to the area $A_{\rm H}$ of the horizon, the low energy Newton constant $G_{\rm N}$ and $\ell$. The `semi-classical input' is the assumption that the physical sector of LQG contains states that are peaked around solutions admitting isolated horizons with such near horizon geometries. Under this assumption, one can use the expression of $E$ and the LQG expression of the area operator $\hat{A}_{\rm H}$ to construct the Hamiltonian operator. In this step, the Planck length $\lp$ in the expression of the area operator is assumed to be given by $\lp^{2} = G \hbar$ where $G$ is the gravitational constant in the Planck regime (which one would expect to run to $G_{\rm N}$ in the low energy regime). Using the Hamiltonian operator (which features $G, G_{\rm N}$ and $\ell$) one can construct a canonical ensemble at the Unruh temperature $T = 2\pi/\ell$ associated with the stationary observers and calculate the entropy. \index{Unruh temperature}  The leading term in the result turns out to be precisely the Bekenstein-Hawking entropy, with no extra factor. The dependence on the auxiliary parameter $\ell$ as well as on the area gap $\Delta$ drops out of this semi-classical result based on the canonical ensemble tailored to stationary, near horizon observers. The calculation also sheds new light on how the possible running of gravitational constant can be naturally accommodated, even when the Hawking-Bekenstein formula uses the low energy value $G_{\rm N}$ of the gravitational constant and the area operator from quantum geometry uses the Planck scale value $G$. \index{Hawking radiation}

Finally, there have been a number of other developments. First, using spin foams, entanglement entropy associated with horizons has been re-examined. It has been shown that if one changes focus from entropy to variation of entropy, a lot of unnecessary conceptual and computational baggage is removed. In particular the species problem is alleviated. It has also been suggested that it is the entanglement entropy that features in a precise statement of the first law in semi-classical gravity and the idea is under more detailed investigation. Another --and major-- direction is a systematic investigation of the quantum geometry underlying a black hole spacetime. This has been successfully carried out under the assumption of spherical symmetry. As in cosmological models, the singularity is naturally resolved by quantum geometry effects. Then it is natural to study quantum fields propagating on this \emph{quantum} black hole geometry. These investigations bring out how the underlying quantum geometry provides a natural ultraviolet regularization in the QFT calculations. Finally, there are calculations of Hawking effect on the \emph{quantum} geometry representing a spherically symmetric black hole, again in the test field approximation. The quantum nature of the underlying geometry leads to ultraviolet modifications of the quantum vacua, eliminating the trans-Planckian modes close to the horizon. As a result, expectation value of the stress energy tensor are ultraviolet finite. Hawking radiation at infinity has been computed and, apart from an ultraviolet cutoff, the outgoing state is the same as the conventional one. In particular, this analysis provides a concrete quantum gravity calculation showing that Hawking radiation is not seriously  contaminated by the apparent trans-Planckian problems of QFT in classical, curved spacetimes. Currently, these calculations are being revisited and extended to study the evaporation process. \index{Hawking radiation}\\


{\it Chapter 8}\, covers a broad range of issues related to phenomenology and observations. They include direct and indirect probes into the very early universe, potential for Lorentz invariance violations, modifications of the spectrum of an evaporating black holes due to the discreteness of area eigenvalues, Planck stars and the possibility that the emission from them is related to Gamma ray bursts (GRBs). In the discussion of cosmological probes there is an inevitable, small overlap with {\it Chapter 6} but the rest of the material is presented for the first time in this volume. The material covered here illustrates the wide scope of current research in LQG, in particular the fact that now there is a significant community that is looking beyond foundational, conceptual and mathematical issues into the interface with observations. 

Conceptually, perhaps the most ambitious of these ideas is that of `Planck stars' which is now superseded by the more recent work on `fireworks'. One knows from quantum cosmology of closed FLRW models that quantum geometry effects dominate and cause a quantum bounce once the curvature reaches the Planck scale. The curvature at the surface of a solar mass collapsing star reaches the Planck scale when its radius is $\sim 13$ orders of magnitude larger than the Planck length. Thus the view that quantum gravity can drastically modify classical dynamics only when the horizon is Planck size is unsubstantiated. It is reasonable to suppose that something like a quantum bounce would occur when the curvature at the surface of the star reaches the Planck scale. The question is: What is the time scale -as measured at infinity- that is involved in this bounce? The time scale for the Hawking evaporation goes as $M^{3}$. If the time scale for the bounce is smaller, say $\sim M^{2}$, then the bounce would dominate, making Hawking evaporation irrelevant for large black holes. The hope in the program is that something like this does happen. This is a bold conjecture. The LQG group pursuing these ideas is well aware of the fact that  extraordinary conjectures require extraordinary proofs and is therefore engaged in building up the necessary evidence through detailed calculations. \index{Planck stars}

Concepts and results that are needed as prerequisites to understand the contents of {\it Chapter 8} have already been discussed in this Introduction. Therefore, we will conclude with just one remark. As in other approaches to quantum gravity, to make contact with observations, one has to make additional assumptions that are not part of LQG proper, since situations of interest to observations are too complex to be systematically arrived at starting from a fundamental quantum gravity theory. Therefore, observations probe the package consisting of the fundamental theory being used in calculations, \emph{together with} the additional assumptions that are made to arrive at phenomenological predictions. This is also the case in other areas of physics, such as QCD. Much of the work on quark-gluon plasmas, for example, makes hydrodynamical assumptions that are physically motivated but are not known to be direct consequences of fundamental QCD. Therefore, if a prediction is falsified, one cannot conclude that there is a problem with QCD; it is \emph{much} more likely that the problem lies with the additional assumptions. Of course, QCD has absolutely huge observational support compared to any quantum gravity theory. But the general spirit of this research in LQG is the same as that in these other areas: use observations to refine paradigms, assumptions and strategies. If a particularly clean prediction is verified, it would give confidence in the underlying assumptions and encourage more detailed analysis within that paradigm, leading to further predictions. If observations contradict a prediction, they provide guidance as to which of the assumptions are suspect and need to be weeded out. However, because observational evidence for any quantum gravity theory is scant, one constantly keeps an eye on the possibility that the observations are suggesting or even requiring a change in the fundamentals of LQG and LQC.  

\section{Closing Remarks}
\label{s6}

Perhaps the most distinguishing feature of LQG is the underlying quantum geometry. Intuitively one expects some sort of discreteness at the Planck scale. Much of the heuristic literature as well as some of the systematic quantum gravity approaches assume that quanta of geometry have a simple structure with lengths that are integral multiples of $\lp$, and/or area of $\lp^{2}$ and/or volume of $\lp^{3}$. The quantum Riemannian geometry of LQG is much more subtle. In particular, the spectra of various geometric operators are quite different from one another; for example, one cannot deduce the eigenvalues of the volume operator from those of the area operator. Secondly, while there is an area gap, higher eigenvalues of the area operator crowd and the level spacing decreases exponentially! Consequently the continuum geometry is approached very quickly. These specific features were first discovered in the 1990s but remained mathematical curiosities until recently. Now such details 
are turning out to be important in the interplay between theory and observations 

For example, in the Bekenstein-Mukhanov approach to black hole entropy, area eigenvalues are of the type $g\, N \lp^{2}$ where $g \sim \mathcal{O}(1)$ can be thought of the degeneracy factor --or, the number of  `bits'-- in the basic quantum of area. This means that when a black hole is perturbed it makes a transition from an initial state of the horizon characterized by an integer $N_{1}$, to a final state characterized by another integer $N_{2}$. Therefore, the change $\Delta A$ in the horizon area occurs in discrete steps: $\Delta A = g\, (N_{2}- N_{1}) \lp^{2}$. Recently, these ideas were re-examined using the relation between area, mass and spin of Kerr black holes, and frequency of quasi-normal modes that describe their `ringing' under perturbations. Specifically, it was argued that the Bekenstein-Mukhanov proposal leads to a different description of ringing and therefore it may soon be under stress once LIGO observes quasi-normal ringing from a sufficient number of spinning black holes. By contrast, for black holes of interest to LIGO, the LQG area spectrum becomes almost continuous due to the exponential crowding of eigenvalues. Therefore, there is no reason for predictions to be  observationally distinct from those of classical general relativity. Another example comes from cosmology, where one can  constrain the value of the area gap $\Delta$ from observations. More precisely, one can leave $\Delta$ as a free parameter in LQC calculations and obtain the value that makes the predicted power spectrum fit best with the PLANCK mission observations. \index{CMB observations} Recent investigations in LQC have carried out this task, providing a completely independent way of arriving at the value of area gap $\Delta$ (and hence of the Barbero-Immirzi parameter $\gamma$). \index{Barbero-Immirzi parameter} The best-fit value of $\Delta$ agrees with that obtained from black hole entropy calculations within the 68\% confidence level used by the PLANCK mission to report its data. LQG researchers who first investigated geometric operators in the 1990s did not anticipate that the details they carefully worked out would start making a difference on observational fronts within just two decades.

For full quantum dynamics, a number of important issues remain. The conceptual framework and mathematical techniques introduced in the early discussions were crucial to get the program rolling. But, with hindsight, we now see that these discussions failed to take into account key issues such as physics at different scales, renormalization group flows, and consistency requirements in the continuum limit procedure. That is why the program has not made as much contact with low energy physics as one would have hoped. Now the community is well aware of these limitations and is actively engaged in overcoming them. {\it Chapters 2-5} summarize the three  directions that are being pursued: Hamiltonian theory a la Dirac, spin foams and their continuum limit, and GFT. While the final goal is essentially the same --uncovering dynamics through solutions to the quantum constraints-- there are also some differences and healthy tensions. The Hamiltonian methods of {\it Chapter 2} focus on canonical gravity while methods used in {\it Chapters 3-5} come from path integrals. They have complementary strengths and limitations. For example, in terms of concrete results in 4 dimensions, the Hamiltonian methods have been much more useful in the study of the early universe and black hole entropy, while path integral methods have shown how one can obtain the graviton 2-point function in Minkowski space in a background independent setting. Even within the path integral approaches there are differences. Roughly, GFT of {\it Chapter 4} is modeled more on methods from QFT, while the continuum limit of spin foams discussed in {\it Chapter 5} generalizes ideas and techniques used in condensed matter physics. 

Given the variety of difficult issues one encounters in non-perturbative quantum dynamics, we believe this diversity is essential. It represents a
key strength of the program. Existing tensions between these sets of ideas will bring to forefront deeper issues and launch new investigations leading to a more coherent overall framework. In particular, they should lead to advances on two key open issues:\\
(i) Contact with the standard \emph{low energy} effective theory. The derivation of the graviton 2-point function, for example, has been a major achievement but one needs to make sure that the current calculations are not changed in the leading order by including spin foams with a large number of vertices, or in an appropriate continuum limit. It is also important to obtain higher order corrections in a reliable fashion; and,\\
(ii) Matter couplings. We know that there is no conceptual difficulty in incorporating matter either in the Hamiltonian framework \cite{aabook,ttbook} or in spin foams \cite{crbook}. But details have not been worked out. In particular, a satisfactory derivation of the classical limit for gravity interacting with matter is still lacking. It is important to understand whether consistency requirements --such as those used in {\it Chapters 2 and 5}-- strongly constrain matter couplings. In these investigations, recent results in the asymptotic safety program may provide guidance.

Finally, results on symmetry reduced models in the study of the early universe and black holes provide important checks on viability of the main ideas underlying LQG. The fact that the program can be completed in these models and that the results successfully address long standing questions of quantum gravity is a non-trivial indication that the program is well-founded. 

However, there is always the issue of whether an important aspect of physics is overlooked by first focusing on a symmetry reduced sector of the classical theory and then passing to the quantum theory. The Dirac model of the hydrogen atom sheds interesting light on this issue. Here one considers the proton-electron system in quantum electrodynamics (QED), truncates the theory to its \emph{spherical symmetric sector} and then carries out quantization. From the perspective of full QED, the strategy seems to introduce a  \emph{drastic} oversimplification since it banishes all the photons right from the start! Indeed, conceptually, the truncated theory does ignore most of the rich possibilities one can envisage in full QED. And yet the Dirac theory provides an excellent description of the hydrogen atom. Indeed, to see its limitations, one has to carry out \emph{very} accurate measurements and carefully examine the hyperfine structure due to QED effects such as the Lamb shift! Returning to quantum gravity, it is not unreasonable to expect that the symmetry reduction strategy would be appropriate \emph{provided} it keeps an eye on the structure of the full theory, i.e., provided the quantum theory of the reduced model is not constructed in a manner that is specifically engineered just to fit that model. In LQG this view is taken seriously: quantum theories of reduced models pay due attention to the quantum geometry that emerged from \emph{full} LQG. Therefore, for the limited class of observables that are needed to describe the very early universe and spherically symmetric black holes, the varied and rich possibilities of full quantum gravity could just refine, rather than alter, the predictions of the reduced models. Of course there is no a priori guarantee that this will be the case. Therefore, over the past 2-3 years there has been an increased effort on finding the precise relation between the symmetry reduced quantum theories and full LQG. In particular, there are strong results showing that the LQC faithfully captures the LQG kinematics in the homogeneous, isotropic sector. At the dynamical level, there are partial results that go in the same direction.\\

We will conclude with a few general remarks about this volume. The list of mathematical symbols that appears in the beginning is intended to help readers as they navigate through detailed arguments. It was sent to all authors and by and large they have followed the conventions laid down in that list. There are several senior figures in LQG who have made invaluable contributions and driven major advances over the past two decades or more. But to provide a fresh perspective that emphasizes future directions, we thought it would be best to invite some of the younger researchers who are currently leading research programs in new directions. We urged them to express their outlook on where we stand and what the strategies are best suited for future advances. Consequently various Chapters express personal visions that sometimes boldly venture beyond the general consensus in the field. This is especially the case on some issues of quantum dynamics and at the interface of theory and observations. In this respect, our intention is captured in the best spirit of the motto: `let a thousand flowers bloom'. The Editors do not subscribe to all the ideas and views expressed in the 8 Chapters. Rather they feel it is important that beginning researchers be guided by younger leaders who will shape the future of LQG. 

\section*{Acknowledgements}
This work was supported by the NSF grants PHY-1505411, PHY-1305000, PHY-16030630, the Eberly research funds of Penn State and the Hearne Institute and the Center for Computation and Technology of Louisiana State University.

\end{document}